\documentstyle{laa}

\begin{document}

\thesaurus{01(10.03.1; 11.01.2; 11.10.1; 11.14.1; 02.01.2; 02.02.1)}
\title{The Galactic Center radio jet}
\author{Heino Falcke \and Karl
Mannheim \and Peter L. Biermann}
\offprints{HFALCKE@mpifr-bonn.mpg.de {\it Preprint Server:} babbage.sissa.it,
 get astro-ph/9308031}
\institute{Max-Planck Institut f\"ur Radioastronomie, Auf den H\"ugel
69, D-53121 Bonn, Germany}
\date{submitted December 23, 1992, accepted August 10, 1993}
\maketitle
\begin{abstract} Recent observations of the
radio and NIR source Sgr~A* reinforce the interpretation of the
Galactic Center as a scaled down version of an AGN.  The discovery of
an elongated structure at 43 GHz and increasing evidence for the
presence of an accretion disk surrounding a Black Hole lead us to
assume that both, an accretion disk and a jet, are present in the
Galactic Center and are physically linked. We model the radio emission
of Sgr~A* successfully with a Blandford \& K\"onigl type jet and
analyze the energetics of the coupled jet-disk system in Sgr~A* where
jet and disk are parametrized in terms of the accretion power.  With
this method we are able to confirm independently the lower limit of
the Sgr~A* accretion rate $\dot{M}\gg10^{-8.5} M_{\sun}$ found
previously.  Moreover, using the limits imposed by observational data,
we show that within such a jet-disk model, the total jet power $Q_{\rm
jet}$ is of comparable order as the radiated disk luminosity $L_{\rm
disk}$. A jet model together with the assumption of an $10^6 M_{\sun}$
Black Hole also qualitatively explains the submm excess and the lack
of non-thermal IR radiation. The small size of the visible part of the
jet $(< 1$ mas) is due to the low accretion rate of Sgr A*.

\end{abstract}

\section{Observations of Sgr~A*}

A prominent phenomenon in the Galactic Center is the compact,
non-thermal radio source Sgr~A* (Balick \& Brown 1974, Lo et al.
1985), showing a fairly flat ( $S_{ \nu } \propto \nu ^{0...-0.3} $)
spectrum in the range 1 to 230 GHz (eg. Zylka \& Mezger 1988).  Eckart
et al. (1992) and Rosa et al. (1992) have detected Sgr~A* at $
\lambda 2.2
\mu
{\rm m}
$ and $
\lambda 1
\mu
{\rm m} $ wavelengths providing {evi\-dence} for a luminous central
object radiating also at optical wavelength. Zylka et al.
(1992) explain their $
\lambda 1300
\mu
{\rm m}
$ and $
\lambda 870
\mu
{\rm m} $ continuum observations of the Sgr~A* region by the existence
of a thermal dust disk surrounding Sgr~A* or a self-absorbed compact
synchroton source. Falcke et al. (1993) explain the NIR data with a
rotating Black Hole of mass $M_{
\bullet }
\sim 10^{6}M_{
\sun }
$ surrounded by a hot accretion disk seen edge-on. And finally, as
another breakthrough, Krichbaum et al. (1993) succeeded in
detecting Sgr~A* with VLBI at 7mm. For the first time a resolved picture
of this non-thermal radio source suggests an elongated structure
on a scale of a few mas corresponding to an observed linear size of several
$10^{14}{\rm cm}$ inclined at $\sim 55^\circ$ to the galactic plane.

\section{The jet-disk coupling}
\subsection{Parametrization of the jet-disk model}
We now want to investigate whether the elongated radio structure in
Sgr~A* can be interpreted as a radio jet. It is commonly believed that
radio jets originate from an accretion disk surrounding a star,
neutron star or a Black Hole. As there are also strong hints for
emission from an accretion disk in Sgr~A* (see above) that assumption
may be valid for the Galactic Center as well, providing us with
 observational information for both systems. We therefore
develop a jet model with regard to the coupling of jet and disk and
the constraints imposed by observation.

Within such a jet-disk model, we express the basic properties of the
radio jet in units of the disk accretion rate $\dot{M}_{\rm disk}$ of
Sgr~A*.  The maximum accretion power $Q_{\rm accr}$ is given by the
rest energy at infinity of the matter $\dot{M}_{\rm disk} c^2$. The
total radiated disk luminosity $L_{\rm disk}$ is an appreciable
fraction $q_{\rm d}$ of this accretion power and in the case of a
Black Hole as the central object will be in the range $q_{\rm d}\sim
5\%-30\%$ (Thorne 1974).  Likewise, the total jet power $Q_{\rm jet}$
-- including the rest energy of the expelled matter -- should be a
fraction $q_{\rm j}<1$ of the accretion power and  also the mass loss rate
due to the jet $\dot{M}_{\rm jet}$ is a fraction $q_{\rm m}<1$ of the
mass accretion rate in the disk.  Thus we define

\begin{equation}
Q_{\rm accr}=\dot{M}_{\rm disk} c^2\!\!,\hfill q_{\rm j}=\frac{Q_{\rm
jet}}{Q_{\rm accr}}, \hfill q_{\rm d}=\frac{L_{\rm disk}}{Q_{\rm
accr}},
\hfill  q_{\rm m} =\frac{\dot{M}_{\rm jet} c^2}{Q_{\rm accr}}. \hfill
\end{equation}
The $q_{\rm d,j,m}$ are dimensionless parameters,
while $Q_{\rm accr}$ defines the physical scale of the system; $c$
denotes the speed of light. We neglect all other energy consuming
processes so that the remaining energy $(1-q_{\rm j}-q_{\rm d}) Q_{\rm
accr}$ is swallowed by the Black Hole.

\subsection{Basic assumptions of the jet-disk model}
The above quantities are all measured in the observer's frame; we will
now switch to the rest frame of the jet and use this parametrization
to calculate the radio emission of Sgr~A* as emission from a radio jet
and analyze its possible link to the accretion process. As the
structural information about the Sgr~A* radio emssion is still
uncertain, we will use the simplified Blandford \& K\"onigl (1979) jet
model which was developed initially for the unresolved core of AGN
jets. The basic idea is that the radio emission is produced by a
supersonic, freely expanding, therefore conical jet with semi-opening
angle $\phi$ and constant velocity $\beta_{\rm j}$, convecting a
tangled magnetic field dominating the internal gas pressure and
producing a powerlaw energy distribution of relativistic electrons via
shock acceleration. It is assumed that there is an approximate
equipartition between the magnetic energy density $u_{\rm
mag}=B^2/8\pi$ and the relativistic particle energy density $u_{\rm
rel}$. In the following we will use cylindrical coordinates, denoting
the axis along the jet as $z$ and the distance from the axis to the
boundary surface of the jet-cone as $r=z\,\sin\phi$. We discuss only
the part of the jet far away from its footpoint.

For this model one usually takes an electron distribution of the form
$N(\gamma_{\rm e})=K \gamma_{\rm e}^{-2}$ for relativistic electron
$\gamma$-factors in the range $\gamma_{\min}\le\gamma_{\rm
e}\le\gamma_{\max}$.  From $u_{\rm rel}\simeq u_{\rm mag}$ we get
$K=B^2/(8\pi m_{\rm e} c^2 \Lambda)$,where $m_{\rm e}$ is the electron
mass and $\Lambda=\ln(\gamma_{\max}/\gamma_{\min})$ yielding a
number density for electrons of $n=B^2/(8\pi\Lambda\gamma_{\min}
m_{\rm e} c^2)$.

The energy density and the pressure in the jet is assumed to be mainly
due to the magnetic field and the relativistic particles $p=p_{\rm
mag}+p_{\rm rel}\simeq (B^2/8\pi) (1+\frac{1}{3})$. Actually the factor
$\frac{1}{3}$ is valid only for the extreme relativistic case where
$p_{\rm rel}=\frac{1}{3}u_{\rm rel}$ and exact equipartition, however,
it can only be as high as $\frac{2}{3}$ in the nonrelativstic case or
negligible for $u_{\rm rel} \ll u_{\rm mag}$. Thus our choice of $p$ is a
fair approximation.

An important parameter of the jet flow is its proper Mach number
${\cal M}={\beta_{\rm j}\gamma_{\rm j}}/{\beta_{\rm s}\gamma_{\rm
s}}$. For a free jet the minimum opening angle is given by its Mach
angle, which is defined as $\sin \phi_{\min}=1/{\cal M}$ (K\"onigl
1980). For a perfect gas we have: $\gamma_{\rm j}=1/\sqrt{1-\beta_{\rm
j}}$, $\beta_{\rm s}={\sqrt{\Gamma p/\omega}}<1$, $\gamma_{\rm
s}\simeq 1$, the adiabatic index $\frac{4}{3}\le\Gamma\le\frac{5}{3}$
, and the enthalpy density $\omega=m_{\rm p} n c^2 + \Gamma
p/(\Gamma-1)$ if we demand equal numbers of thermal protons and
relativistic electrons (cnf. last paragraph of this section).  Here we
expect $\Lambda\gamma_{\min}\ll\frac{m_{\rm p}}{m_{\rm e}}$ and thus

\begin{equation}\label{machnumber}
{\cal M}={\gamma_{\rm j} \beta_{\rm j}}/\sqrt{{\frac{4 m_{\rm
e}}{3 m_{\rm p}}\Gamma \Lambda\gamma_{\min}}},
\end{equation}
for the other extreme we would get a maximum sound velocity $\beta_{\rm
s,\max}=\sqrt{(\Gamma-1)}$ and the condition ${\cal M}>\gamma_{\rm j}
\beta_{\rm j}/\beta_{\rm s,\max}$.

Considering that the particle flux of the jet is given by $q_{\rm m}
Q_{\rm accr}/m_{\rm p} c^2=\gamma_{\rm j} n \beta_{\rm j} c \pi r^2$ we can now
express the magnetic field in the jet in terms of the numbers $q_{\rm
m}$ and ${\cal M}$, obtaining the familiar $B\propto z^{-1}$ behaviour.

\begin{equation}
B=\sqrt{\frac{6 \gamma_{\rm j} \beta_{\rm j} q_{\rm m} Q_{\rm accr}}{\Gamma
{\cal M}^2 z^2 \sin\phi^2 c}},
\end{equation}

The assumption of Blandford \& K\"onigl that all electrons are
relativistic and all protons are thermal is very crude and we would
rather expect to have a mixture of a thermal and a relativistic plasma
both containing electrons and protons. But if the ratio of thermal
protons to nonthermal electrons is constant throughout the jet, this
effect will cancel out in the final equations and only the definition
of $\Lambda$ and ${\cal M}$ in Eq. (\ref{machnumber}) will
change.  We therefore will not use Eq. (2) in this {\it Letter} and rather
consider $\cal M$ and $\Lambda$ as free parameter which in principle
contain all the unknown microphysics (e.g. the electron to proton
ratio). We leave a more elaborated discussion of this point to a
subsequent paper.

\subsection{Energy equation of the the jet-disk model}
To investigate how much energy is extracted by the jet relative to the
disk, we have to write down the basic energy equation for the
jet-disk system. This is simply the relativistic Bernoulli equation for
the jet $\gamma\omega/n=\left.\omega/n\right|_\infty$ with the left
hand side parametrized by the quantities $q_{\rm m}$ and ${\cal M}$ and
the right hand side determined by the energy supply from the
accretion process and parametrized by $q_{\rm j}$. Algebraic
transformations then lead to
\begin{equation}\label{energy}
\gamma_{\rm j} q_{\rm m} \left(1+\frac{\gamma_{\rm
j}^2\beta_{\rm j}^2}{(\Gamma-1){\cal M}^2}\right)=q_{\rm j}.
\end{equation}

If we subtract the rest energy from both sides and take the
non-relativistic limit, this equation simply states that the total power
in the jet consists of the power in relativistic particles and the
magnetic field plus the kinetic energy of the matter.

\begin{figure*}
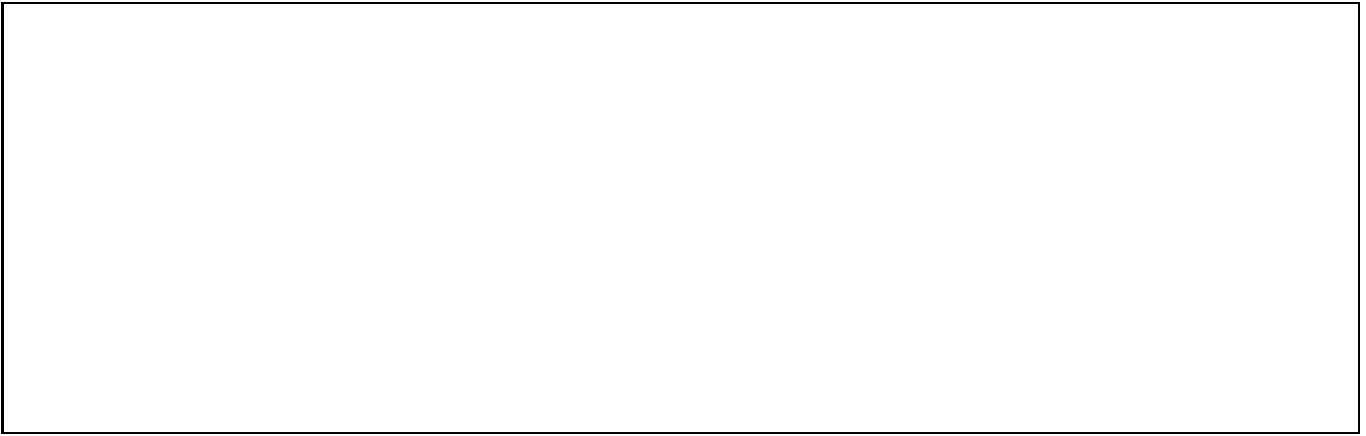

\picplace{5.7cm}
\caption[]{
The shaded area limits the parameter space for a coupled jet-disk
system in Sgr~A* producing a flat radio spectrum with $F_\nu=1.1 Jy$
and a NIR-UV disk luminosity $\la 5\cdot10^5 L_{\sun}$.  The vertical axis
gives the log of the total power extracted by the jet from the
accretion disk normalized by the total accretion power $Q_{\rm
accr}=\dot{M}_{\rm disk}c^2$; for comparison: the radiative power of
the accretion disk is $0.05-0.3 Q_{\rm accr}$.  The horizontal axis
gives the log of the relative mass loss due to the jet compared to the
mass accretion rate of the disk.  Strongly bended lines in the plot
represent models of equal Mach number ${\cal M}$ whereas the less bended
lines represent models of constant proper jet velocity $\gamma_{\rm
jet}\beta_{\rm jet}$. The left figure is plotted for a jet seen side
on with inclination angle $i=90^{\circ}$, while the figure on the
right is plotted for a jet with an axis inclined to the line of sight
by $30^{\circ}$. A shift of the parameter grid with different
accretion rates is indicated by text boxes and the shape of the
shifted bounding lines of the parameter grid.}
\end{figure*}

\subsection{Radio emission from the jet}
To deduce some of the parameters of the jet-disk system from the
observation we now have to calculate the synchroton emission from the
jet described above. Angle averaged emissivity and absorption
coefficient are given by $\epsilon_{\rm sync}=2450\, (B/{\rm
Gauss})^{7/2}(\nu/{\rm GHz})^{-1/2} \Lambda^{-1} {\rm Jy}/{\rm cm\,
ster}$ and $\kappa_{\rm sync}=2.25 \cdot10^{-12} (B/{\rm Gauss})^4
(\nu/{\rm GHz})^{-3} \Lambda^{-1} {\rm cm}^{-1}$ respectively. At a
given frequency the jet will become optically thin and thus visible
for a comoving observer at a jet position $z_{\tau_{\nu}=1}$ where the
optical depth along the line of sight, inclined by the angle $i$ to
the jet axis, is unity. For a ray right through the central axis of the
jet and $i\gg\phi$ we can approximate this by
$2\sin\phi\,z_{\tau_{\nu}=1}\,\kappa_{\rm sync}(z_{\tau_{\nu}=1})
/\sin i\simeq1$ neglecting the gradient in the magnetic field along
the line of sight for $i\neq\pi/2$.  To obtain the total flux at this
frequency we then have to integrate the jet emission from
$z_{\tau_{\nu}=1}$ to infinity (or whatever one considers to be the
outer edge of the jet) yielding a flat spectrum. Transforming
the equation into the observer's frame, we have for the continuous,
optically thin part of a jet (Lind
\& Blandford 1985)
\begin{eqnarray}
&&{\cal D}=1/\gamma_{\rm j}(1-\beta_{\rm j}\cos i_{\rm
obs}),\;\;\;F_{\rm obs}(\nu_{\rm obs})={\cal D}^2 F_{\rm obs}({\cal
D}\nu)\nonumber\\ &&\nu=\nu_{\rm obs}/{\cal D},\;\;\;\sin i= {\cal D}
\sin i_{\rm obs},\;\;\; \phi=\phi_{\rm obs} \sin i_{\rm
obs}\end{eqnarray}
and obtain as the observed flux from one jet cone
\begin{eqnarray}\label{Fnu}
F_{\rm obs,1|2}(\nu_{\rm obs})&=&1\,\,{\rm Jy}\cdot\,{\cal
D}^{13/6}{{\sin { i^{{1/ 6}}}}}\, \left({{\cal M}\over 3
}\right)^{-11/6} {{\left({\Lambda\over9}\right)}^{-{5/6}}}\nonumber\\
&&\cdot \left(\gamma_{\rm j}\beta_{\rm j}\,{q_{\rm m}\over
3\%}\,{\dot{M}_{\rm disk}\over 10^{-7}M_{\odot}/{\rm
yr}}\right)^{17/12}
\end{eqnarray} which is the  original Blandford \& K\"onigl
equation only with a new parametrization and the
assumption that the opening angle of the jet is given by the Mach
angle. We adopted a Galactic Center distance of $D=d_{8.5} 8.5$ kpc
and a central accretion rate of $\dot{M}_{\rm disk}=\dot{m}_{-7}
10^{-7} M_{\sun}/{\rm yr}$.  The total flux from both
cones is $F_{\rm obs}=\left.F_{\rm obs,1}\right|_{i_{\rm
obs}}+\left.F_{\rm obs,2}\right|_{\pi-i_{\rm obs}}$.

\subsection{Size of the emitting region}
To find out whether the central core of the jet can be resolved, we
transform the source size in the rest frame given approximately by
$z_{\tau_{\nu}=1}$ into the observed size $z_{\tau_{\nu}=1,\rm obs}$
which is

\begin{samepage}
\begin{eqnarray}\label{size}
z_{\tau_{\nu}=1,\rm obs}&=&2\cdot 10^{13} {\rm cm} \left({{43
GHz}\over{\nu}}\right)\nonumber\\ &&\cdot\left({\gamma_{\rm
j}\beta_{\rm j}
\over\sqrt{1+\left(\gamma_{\rm j}\beta_{\rm j}\right)^2}}
\sqrt{{9\over\Lambda} {3\over{\cal M}}} \,{q_{\rm m}\over
3\%}\,{{\dot{M}_{\rm disk}}\over{10^{-7}M_{\odot}/{\rm
yr}}}\right)^{2/3}\!\!\!\!\!\!\!\!\!.
\end{eqnarray}
\end{samepage}

We see that even at high frequencies the size of the jet is smaller
than current VLBI resolution. If we go even further, i.e. to submm
wavelengths, the size of the emitting region approaches the minimal
size, which is the size of a Black Hole of mass $\sim 10^6 M_{\sun}$.
This scale then should correspond to the `nozzle' of the jet,
producing the highest frequencies and fluxes and explaining the steep
rise of the spectrum in the submm range (Zylka et al. 1992) as a
self-absorbed, very compact and thus variable synchroton source at the
footpoint of the jet (and not as thermal dust emission). At higher
frequencies (IR) non-thermal jet emission should not be detectable
because of the large size of the Black Hole (Falcke \& Biermann 1993).

\subsection{Parameter space of jet-disk models}
Now we can limit the parameter space for possible models of the
jet-disk system in the Galactic Center. We know from observation
that the flux from Sgr~A* at 43 GHz is $F_{\rm obs}\simeq 1.1$
Jy (Krichbaum et al. 1993). Inserting this into equation (\ref{Fnu})
we get an equation for the relative mass loss $q_{\rm m}$ in Sgr~A* as
a function of the opening angle, the velocity, the inclination angle
and the Mach number of the jet. With $q_{\rm m}$ specified, we can
calculate the energy consumption $q_{\rm j}$ of the jet relative to
the disk from the energy equation (\ref{energy}). As a free jet should
have an opening angle $\sin\phi\ga 1/{\cal M}$ and the radiative
efficiency of the jet decreases with increasing opening angle the
equality $\sin\phi=1/{\cal M}$ used in Eq. (\ref{Fnu}) gives a lower
limit for the energy demand $q_{\rm j}$ of the Sgr~A* jet. Moreover we
are not free to choose the other parameters completely arbitrarily.
For a free jet the Mach number should be greater than one and as
stated above for a given Mach number the jet velocity can not exceed
$\gamma_{\rm j,max}\beta_{\rm j,max}={\cal
M}\sqrt{(\Gamma-1)}$. Because of mass conservation in the
jet-disk system we have $q_{\rm m}<1$ and from energy conservation we
get the limit $q_{\rm j}<1$. The latter is probably too
generous, as there is at least some energy loss through the disk
radiation and could be reduced to $2/3$ or even $1/3$ -- the maximum
efficieny of the disk.

Lastly, we need to specify the accretion rate of Sgr~A*. Falcke et al.  (1993)
deduced a range of $10^{-7} M_{\sun}/{\rm yr}>\dot{M}>10^{-8.5}
M_{\sun}/{\rm yr}$ from standard accretion disk models and IR/NIR
luminosities. The upper limit is obtained by assuming that most
of the disk luminosity is absorbed in the surrounding dust and
thus the disk luminosity can not exceed the dust luminosity. But as
the dust is probably also heated by stars of the central cluster, the
real disk luminosity will be much lower.  On the other hand these
models did not include the influence of a jet on the disk structure
and efficiency.  This may change the estimate of the
accretion rate if $q_{\rm j}$ is comparable to the radiative
efficiency of the disk $q_{\rm l}$.

\section{Results}
In Figure 1 we have plotted the lower limit of the relative energy
demand $q_{\rm j}$ and the relative mass loss rate $q_{\rm m}$ of the
Sgr~A* jet compared to the accretion power and mass accretion rate of
the disk as functions of the jet velocity and Mach number for two
inclination angles and different accretion rates. The shaded area
limits the possible parameter space for a Sgr~A* jet-disk system being
capable to produce 1.1 Jy radio emission. For our calculations we took
as adiabatic index $\Gamma=4/3$ and a fairly low logarithmic factor
$\Lambda=9$ which corresponds to a ratio of
$\gamma_{\max}/\gamma_{\min}\sim 10^4$. These values are completely
arbitrary, but a higher $\Lambda$ would again reduce the radiative
efficiency, increase the energy demand and thus strengthen our
conclusion. Note that the models with extremely low Mach number ${\cal
M}\la3$ require unrealistic large opening angles and the lowest part of
the parameter grid is only plotted for completeness.

One can see that even for the highest disk accretion rates allowed by
the IR dust observations the {\em minimum} energy requirements for a
Blandford \& K\"onigl type jet are at least several per cent of the
total accreted rest mass energy $q_{\rm j}>2$\%. If one compares this
with the radiative disk efficiency $q_{\rm l}\sim5\%-30\%$ one can say
that within this model the Sgr~A* jet could extract as much energy
from the disk -- in form of kinetic and magnetic energy -- as is
radiated by dissipative processes in the disk itself.

The minimum energy demands for the Sgr~A* jet are found for low Mach
numbers -- this means high magnetic energy compared to the kinetic
energy -- and moderate relativistic velocities with $\gamma_{\rm
j}\beta_{\rm j}\sim0.1-6$. The Machnumber is limited to be ${\cal
M}<40$. Higher $\gamma$-factors or nonrelativistic velocities seem to
be excluded.  This is well within the range of parameters which is
found in extragalactic jets.

On the other hand our analysis also gives a strict lower limit for the
Sgr~A* accretion rate namely $\dot{M}\gg10^{-8.5} M_{\sun}/{\rm yr}$
which is independent of the disk luminosity. Any lower accretion rates
are not capable of producing such a strong radio emission. This
agrees well with the lower limit given by Falcke et al. (1993) deduced
earlier. And finally it should be noted that if one accepts the high
energy extraction efficiency of the jet -- despite our ignorance of
the exact physical process being responsible for this -- there is
ample space in the parameter space left to positively state that the
radio spectrum of Sgr~A* can be understood as emission from a radio
jet which is barely resolved due to a small spatial scale and a small
power, both limited by the low disk accretion rate.

\section{Discussion}
Together with the spectra discussed in preceding papers (Zylka et al.
1992, Falcke et al. 1993) we can explain the spectrum of Sgr~A* from
radio to NIR in a consistent manner, including the flat radio spectrum
(jet), the upturn in the submm range (`nozzle'), the break in the IR
(Black Hole) and a new component in the NIR (disk). It shows that such
an AGN like scenario with Black Hole, accretion disk and jet may
explain the inner part of the Galactic Center as well, but on a much
lower level of activity.  This is caused by a very low central
accretion rate in the range $10^{-7} M_{\sun}/{\rm
yr}>\dot{M}>10^{-8.5} M_{\sun}/{\rm yr}$ making the Galactic Center an
AGN on a starvation diet.

The elongated structure found by Krichbaum et al. (1993) on the mas
scale is well explained as a weak radio jet. On the kpc scale Sofue et
al. (1989) found a large structure (`Galactic Center Spur') pointing
at the Galactic Center which could well be the final smoke trail of
our Galactic Center jet.

Our finding that the energy extracted by the jet in Sgr~A* is
comparable to the energy radiated by the accretion disk is not unique.
Rawlings \& Saunders (1991) came to a similar conclusions
from a purely observational analysis of jet and disk emission of a
large variety of radio galaxies. What is striking here, is that these
radio galaxies are ususally steep-spectrum radio sources associated
with an active galactic nucleus inside elliptical galaxies. If indeed
an ordinary spiral like our Milky Way shows a similiar $Q_{\rm
jet}/L_{\rm disk}$ ratio as these AGN, than the ratio of jet power to
disk luminosity is a consequence of fundamental jet-disk physics and
not special to any type of galaxy, whereas the absolute powers of jets
and disks are a consequence of different host galaxies, i.e. the
different processes of feeding. Interestingly our equation (\ref{Fnu})
predicts a certain scaling of radio emission from the central
radio core with accretion rate which should hold for AGN as well.

The combined jet-disk system we have sketched here, is of course very
simplified, but as we have concentrated on the basic energetics this
seems to be justified. Moreover, the statement that the jet extracts a
non negligible part of the accretion power is fairly robust, as we
always used lower limits for the energy demand of the jet.

Nevertheless, some assumptions of the Blandford \& K\"onigl model may
be violated but only a few of them making a lower $Q_{\rm jet}/L_{\rm
disk}$ possible. For example the jet could be additionally confined
with an opening angle smaller than the Mach angle, and therefore be
more efficient. Perhaps with better data at hand (e.g. velocity and
opening angle of the jet or the mass accretion rate) such an
additional confinement may even be required to explain the high
efficiency of the Sgr~A* radio jet compared to the luminosity of the
accretion disk. On the other hand, the jet could extract additional
rotational energy from the Black Hole (Blandford \& Znajek 1977) or
reduce the disk luminosity by its influence on the accretion process,
which we have not included in our model. There are also spherical
accretion scenarios (Melia et al. 1992) which, however, seem to be
disproven by the observation of a non-spherical structure.  Finally
there could be a less efficient central object -- other than a Black
Hole -- or a not fully dissipative disk changing the estimates for the
accretion rate.

\acknowledgements
HF is supported by the DFG (Bi 191/9).  KM acknowledges support by
DARA grant FKZ 50 OR 9202.  We thank W. Duschl, T. Krichbaum, J.
Rachen, R. Zylka and two of the referees for helpful discussions.
 \end{document}